# Weak Lensing by Large Scale Structure in Open, Flat, and Closed Universes

**Jens Verner Villumsen**
*Max Planck Institut für Astrophysik,*
*Karl Schwarzschild-Str. 1, Postfach 15 23, 85740 Garching, Germany*
*jens@mpa-garching.mpg.de*

2 March 1995

**ABSTRACT**
Weak lensing is the distortion (polarization) of images of distant objects, such as high redshift galaxies, by gravitational fields in the limit where the distortion is small. Gravitational potential fluctuations due to large scale structure cause correlated distortions of the images of high redshift galaxies. These distortions are observable with current large telescopes and instrumentation.

In a Friedmann-Robertson-Walker (FRW) metric I calculate the weak lensing pattern due to large scale structure for arbitrary $\Omega_0$ and zero cosmological constant $\Lambda$. For a given cosmological model, specified by $\Omega_0$ and a power spectrum of density fluctuations, I calculate the statistical properties of the polarization field for an arbitrary redshift source distribution in a simple closed form.

It is shown that for low redshift $z$ of the sources, the polarization amplitude is proportional to $\Omega_0$, while at higher redshift the polarization measures the value of $\Omega(z)$, where $z$ is the characteristic source redshift. Moreover, the statistics of the polarization field are a direct measure of the power spectrum of density fluctuations.

# 1   INTRODUCTION

Gravitational lensing provides a direct probe of mass fluctuations in the Universe. In strong lensing by individual galaxies or clusters there is a significant distortion of images even to the point of forming multiple images. This was first proposed by Zwicky 1937 and was first observed by Walsh, Carswell & Weyman 1979. Since then, a large number of strong lensing systems have been discovered (see f.ex. Schneider, Ehlers & Falco 1992). Weak lensing, on the other hand, is generated by sufficiently weak gravitational fields and manifests itself in a distortion of images of distant objects, such as high redshift galaxies. A distant galaxy seen behind a high density region will be elongated tangentially relative to the line of sight to the cluster center. The same galaxy will be elongated radially relative to the void center if the light rays pass through a void. As the light from a galaxy travels through high and low density regions towards the observer the galaxy image will be distorted in random directions.

Weak lensing is therefore a measure of the amplitude and coherence of the cosmological gravitational field and as such is a measure of the mass distribution in the universe. The important point about weak lensing is that although on average the image distortions will cancel out, within a given small patch of the sky it will induce a preferred direction.

We parametrize galaxy shapes by the complex eccentricity $\chi = -\epsilon \exp(2i\phi)$ where $\epsilon$ is the eccentricity of the image and $\phi$ is the position angle of the major axis. We make the assumption that in this patch of the sky there is intrinsically no preferred direction, i.e. $<\chi_{true}>=0$. The polarization $p$ is a measure of the induced complex eccentricity of an image. If we have an image with true complex eccentricity $\chi_{true}$, the observed complex eccentricity $\chi_{obs}$ will be

$$\chi_{obs} = \chi_{true} + p\left(1 - \frac{1}{2}\chi_{true} \cdot \chi_{true}^*\right) - \frac{1}{2}p^*\chi_{true}^2 \tag{1}$$

where $*$ denotes complex conjugate. In the presence of weak lensing, the observed mean value of the complex orientation $<\chi>$ is a measure of the polarization $p$ due to weak lensing. $<\chi>\approx p$. The polarization due to weak lensing is thus an observable quantity.

In an individual cluster the polarization field carries information about the depth and shape of the gravitational potential. Weak lensing in clusters has been detected in a large number of systems, e.g. Tyson, Valdes & Wenk 1990, Kaiser & Squires 1993, Seitz & Schneider 1994, Smail *et al.* 1994.

If we focus on the polarization due to large scale structure rather than by individual objects, it can be shown that the statistical properties of this polarization field carries information about cosmological parameters such as the value of the cosmological density parameter $\Omega_0$ and the power spectrum of density fluctuations $P(k)$. The first suggestion that weak lensing could be used to probe the large scale mass distribution was made by Gunn 1967 and the first observational investigation by Kristian 1967. Later fairly shallow searches by e.g. Valdes, Tyson & Jarvis 1983 were also unsuccesful. Recent surveys, e.g. Mould *et al.* 1994 (M94), have yielded marginal detections. These are small, very deep pilot surveys. Their importance is that they demonstrated that it is possible to control the systematic errors to the point that larger surveys under excellent atmospheric conditions can detect a signal. In particular the polarization signal in the Sloan Survey (Gunn & Knapp 1994) will offer the opportunity to map in this way the gravitational field of the Local Supercluster, as has recently been suggested by Gould & Villumsen (1994).

Weak lensing calculations have been performed by a number of authors, e.g. Blandford *et al.* 1991 (B91), Miralda-Escude 1991, Kaiser 1992, M94. For a universe with $\Omega_0 = 1$ and a density contrast of approximately unity on a scale of $800 \text{kms}^{-1}$, they predict a mean polarization of a few percent in a deep field of a few square arcmins. This is observable in a modest observational program on a large telescope under excellent atmospheric conditions.

In this paper I calculate the basic statistical properties of the polarization field in a universe with an arbitrary value of $\Omega_0$, assuming $\Lambda = 0$. Instrumental in this calculation is the use of the comoving angular diameter distance as the fundamental variable, which makes it possible to extend previous calculations for Einstein-de Sitter universes in deriving closed form expressions for the various relevant quantities. The calculation is for an arbitrary source redshift distribution, whereby it is assumed that this distribution is the same all over the sky. To be able to perform these calculations I have to introduce a few technical approximations. Firstly, I use the Born-approximation, which says that the distortion of high redshift galaxy images can be calculated as an integral along the geodesic in a universe with no density fluctuations. In addition, it is also necessary to assume the plane-parallel approximation. The latter involves the assumption that the only waves that contribute significantly to the lensing are the ones nearly perpendicular to the line-of-sight, if the density field is decomposed into plane waves. Finally, also for $\Omega_0 \neq 1$ Universes I assume I can carry out a Fourier analysis. Such an analysis is strictly valid only for wavelengths much smaller than the radius of curvature of the universe, and therefore contains the implicit assumption that the contribution of horizon-scale perturbations is negligible. In practice this puts constraints on the power spectrum behaviour at low wave numbers. In general, a power spectrum that runs like $P(k) \propto k^n$, with $n > 0$, at small $k$ will behave well. Formally the polarization will be finite for $n > -2$ but the assumptions in this paper will no longer be valid on large scales.

Some of the present paper has been covered previously by B91, Miralda-Escude 1991, Kaiser 1992, M94. However, they restrict themselves to the simpler problem of an $\Omega_0 = 1$ Universe. I show that the most important cosmological effect is that at low redshift the polarization is nearly proportional to $\Omega_0$. At higher redshift the polarization amplitude is nearly proportional to the value of the cosmological density parameter at the characteristic redshift of the sources. Polarization measurements are therefore direct measurements of $\Omega_0$. Moreover, the dependence of the polarization on the redshift $z$ might contain interesting information on $\Lambda$. The equivalent calculation for auniverse with a non-zero cosmological constant will be dealt with in a future paper.

An introduction to weak lensing by large scale structure is given in §1. This if followed in §2 by the calculation of the polarization field on the sky. The statistical properties of the polarization field are derived in §3. A summary of the equations for the polarization field and the basic statististical properties is given in §4, while in §5 I finish with a short discussion of the results.

## 2 CALCULATING THE POLARIZATION FIELD

### 2.1 Basics

I want to calculate the evolution of the the magnification and shear (distortion of images of high redshift galaxies) in an FRW Universe which is not necessarily flat, but that has $\Lambda = 0$. I use the Friedmann equations for the evolution of the expansion factor $a$, the cosmological density parameter $\Omega(a)$, the Hubble parameter $H(a)$, the comoving angular diameter distance $x$ and the affine parameter $\lambda$. The normalisation is: $a = 1$ is the present time, $t = t_0$, $H_0 = H(a = 1) = 1$, $\Omega(a = 1) = \Omega_0$. The speed of light is $c = 1$. The absolute value of the Hubble constant does not explicitly enter into the results. All the equations are written with $x$, the comoving angular diameter distance as the independent variable. $x$ will henceforth be referred to simply as "distance".

The derivations are an extension and generalisation of B91 and M94 for a weakly perturbed FRW Universe. It is not necessary to make simplifying assumptions beyond those presented for a flat universe. We define a distance polar $D = (D_1, D_2)$, which is a generalised proper distance. Both components of $D$ are in general complex numbers. The real part of $D_1$ represents the distance while the imaginary part measures the image rotation. As shown in B91 the image rotation is negligible and $D_1$ will be treated as a real number. $D_2$ measures the complex rate of shear. In a homogenous universe $D_1$ is the angular diameter distance $D_{OS}$, while $D_2 = 0$. We can write the magnification $\Delta M$ and polarization $p$ in terms of $D$.

$$\Delta M = 2\left(1 - \frac{D_{1r}}{D_{OS}}\right) \; ; \; p = 2\frac{D_2}{D_1} \tag{2}$$

These quantities are first order approximations.

The fundamental equations for the distance polar $D$ are given by B91 and by Seitz, Schneider, & Ehlers 1994.

$$\frac{d^2 \delta D_1}{d\lambda^2} - \mathcal{R}\delta D_1 = \delta \mathcal{R} D_1 \tag{3}$$

$$\frac{d^2 D_2}{d\lambda^2} - \mathcal{R} D_2 = \mathcal{F} D_1 \tag{4}$$

Here $\mathcal{R}$ and $\mathcal{F}$ are derived from the Ricci tensor and the Weyl tensor. In this approximation they can be represented by second derivatives in comoving coordinates of the comoving Newtonian gravitational potential $\Phi$.

$$\mathcal{R} = -a^{-5}\left(\frac{\partial^2 \Phi}{\partial \xi^2} + \frac{\partial^2 \Phi}{\partial \eta^2}\right) \tag{5}$$

$$\mathcal{F} = -a^{-5}\left(\frac{\partial^2 \Phi}{\partial \xi^2} - \frac{\partial^2 \Phi}{\partial \eta^2} + 2i\frac{\partial^2 \Phi}{\partial \xi \partial \eta}\right) \tag{6}$$

Here $(\xi, \eta)$ are coordinates transverse to the direction of the ray. In the plane parallel approximation we can add the second derivative of the potential along the ray.

$$\mathcal{R} = -a^{-5}\nabla^2\Phi \tag{7}$$

The quantity $\delta\mathcal{R}$ is the perturbation of $\mathcal{R}$ relative to a homogeneous universe. Because there is no gravitational shear in a homogeneous univers, $\delta\mathcal{F} = \mathcal{F}$. In these equations $D_1$ will be real, so the image rotation will be zero. We set $D_1$ in Eqs.( 3, 4) to be the unperturbed distance so

$$D_1 = ax \; ; \; \delta D_1 = a\delta x_1 \; ; \; D_2 = a\delta x_2 \; ; \; \Delta M = -2\frac{\delta x_1}{x} \; ; \; p = 2\frac{\delta x_2}{x} \tag{8}$$

Here $(\delta x_1, \delta x_2)$ is the longitudinal and transverse displacement of the beam relative to the beam position in an unperturbed universe. Notice that $\delta x_1$ is a complex coordinate. We can then rewrite the fundamental equations in terms of $x$, the coordinate along the unperturbed ray.

$$\frac{d^2(a\delta x_1)}{d\lambda^2} - \mathcal{R}a\delta x_1 = \delta\mathcal{R}ax \tag{9}$$

$$\frac{d^2(a\delta x_2)}{d\lambda^2} - \mathcal{R}a\delta x_2 = \mathcal{F}ax \tag{10}$$

$$\mathcal{R} = -a^{-5}\nabla^2\Phi = -\frac{3}{2}\frac{\Omega H^2}{a^2} = -\frac{3}{2}\frac{\Omega_0}{a^5} \tag{11}$$

$$\delta\mathcal{R} = -\frac{3}{2}\Omega_0 a^{-5}\delta(\underline{x}) \tag{12}$$

$$\mathcal{F}(\underline{x}) = -a^{-5}\left(\frac{\partial^2\Phi}{\partial\xi^2} - \frac{\partial^2\Phi}{\partial\eta^2} + 2i\frac{\partial^2\Phi}{\partial\xi\partial\eta}\right) \equiv -\frac{3}{2}\Omega_0 a^{-5}\mathcal{F}'(\underline{x}) \tag{13}$$

$\delta(\underline{x})$ and $\mathcal{F}'$ are the density contrast and the shear in the comoving gravitational potential at comoving position $\underline{x}$ at the appropriate epoch $a = a(x)$. The source of the magnification is the density fluctuations $\delta(\underline{x})$ while the source of the shear of the images is the shear $\mathcal{F}$ in the comoving potential.

$\mathcal{F}$ is written in this way for notational consistency with $\delta\mathcal{R}$. For a flat universe both $\delta(\underline{x})$ and $\mathcal{F}'(\underline{x})$ are proportional to the expansion factor $a$ in linear theory. A set of useful relations for the calculation follows.

$$\lambda(t) = \int_t^{t_0} a(t')dt' \Rightarrow \lambda(a) = \int_a^1 \frac{da'}{H(a')} \tag{14}$$

$$x(t) = \int_t^{t_0} \frac{dt'}{a(t')} \Rightarrow x(a) = \int_a^1 \frac{da'}{H(a')a'^2} \tag{15}$$

$$\frac{dx}{da} = \frac{-1}{H(a)a^2} \; ; \; \frac{d\lambda}{da} = \frac{-1}{H(a)} \; ; \; \frac{dx}{d\lambda} = \frac{1}{a^2}; \; \frac{d^2x}{d\lambda^2} = 2\frac{H(a)}{a^3} \tag{16}$$

$$H(a) = a^{-1}\left(1 + \Omega_0(1/a - 1)\right)^{1/2}; \; \frac{\Omega_0}{\Omega(a)} = \Omega_0 + (1 - \Omega_0)a \tag{17}$$

$$\frac{d^2a}{dx^2} = \frac{\Omega_0}{2} + (1 - \Omega_0)\,a; \; \Omega(a)H^2(a) = \frac{\Omega_0}{a^3} \tag{18}$$

### 2.2 Calculation for Single Source

Let me rephrase our standard equation for a source at distance $x$. I will only show the calculation of $\delta x_1$ since the calculation for $\delta x_2$ is nearly identical.

We first change the dependent variable from the affine parameter $\lambda$ to the distance $x$. The we reduce as much as possible the occurences of the expansion factor $a$ and functions of $a$.

$$\frac{d^2(a\delta x_1)}{dx^2}\left(\frac{dx}{d\lambda}\right)^2 + \frac{d(a\delta x_1)}{dx}\frac{d^2x}{d\lambda^2} + \frac{3}{2}\Omega_0 a^{-4}\delta x_1 = -\frac{3}{2}\Omega_0 a^{-4}\delta(\underline{x})x \quad\Rightarrow$$

$$\frac{d^2(a\delta x_1)}{dx^2}a^{-4} + 2\frac{d(a\delta x_1)}{dx}a^{-3}H(a) + \frac{3}{2}\Omega_0 a^{-4}\delta x_1 = -\frac{3}{2}\Omega_0 a^{-4}\delta(\underline{x})x \quad\Rightarrow$$

$$\frac{d^2(a\delta x_1)}{dx^2} + 2\frac{d(a\delta x_1)}{dx}aH(a) + \frac{3}{2}\Omega_0\delta x_1 = -\frac{3}{2}\Omega_0\delta(\underline{x})x \quad\Rightarrow$$

$$a\frac{d^2\delta x_1}{dx^2} - 2H(a)a^2\frac{d\delta x_1}{dx} + \left(\frac{\Omega_0}{2} + (1 - \Omega_0)\,a\right)\delta x_1 +$$

$$2\left(-H(a)a^2\delta x_1 + a\frac{d\delta x_1}{dx}\right)aH(a) + \frac{3}{2}\Omega_0\delta x_1 = -\frac{3}{2}\Omega_0\delta(\underline{x})x \quad\Rightarrow$$

$$\frac{d^2\delta x_1}{dx^2} + (\Omega_0 - 1)\,\delta x_1 = -\frac{3}{2}\Omega_0\frac{\delta(\underline{x})}{a}x \tag{19}$$

Equivalently the equation for the shear is

$$\frac{d^2\delta x_2}{dx^2} + (\Omega_0 - 1)\,\delta x_2 = -\frac{3}{2}\Omega_0\frac{\mathcal{F}'(\underline{x})}{a}x \tag{20}$$

The boundary conditions are

$$\delta x_1 = 0 \; ; \; \frac{d\delta x_1}{dx} = 0 \; ; \delta x_2 = 0 \; ; \; \frac{d\delta x_2}{dx} = 0 \; ; \text{for } x = 0 \tag{21}$$

We have not been completely successful in getting rid of $a$, we are left with one power of $a$. It is, however, not profitable to express $a$ in terms of $x$ and $\Omega_0$ in these equations. We also see that the solutions to the homogeneous equations change character depending on whether the universe if open, flat, or closed. The solutions are exponential, linear, or trigonometric functions depending on the value of $\Omega_0$.

With these boundary conditions the general solutions are

$$\begin{aligned}
\frac{2\delta x_1}{x} &= -3\Omega_0 \int_0^x \frac{\delta(\underline{x}')}{a(x')} \frac{x'(x-x')}{x} \frac{\sin\left[(\Omega_0-1)^{1/2}(x-x')\right]}{(\Omega_0-1)^{1/2}(x-x')} dx' \quad \text{for } \Omega_0 > 1 \\
&= -3 \int_0^x \frac{\delta(\underline{x}')}{a(x')} \frac{x'(x-x')}{x} dx' \quad \text{for } \Omega_0 = 1 \\
&= -3\Omega_0 \int_0^x \frac{\delta(\underline{x}')}{a(x')} \frac{x'(x-x')}{x} \frac{\sinh\left[(1-\Omega_0)^{1/2}(x-x')\right]}{(1-\Omega_0)^{1/2}(x-x')} dx' \quad \text{for } \Omega_0 < 1
\end{aligned} \tag{22}$$

$$\begin{aligned}
\frac{2\delta x_2}{x} &= -3\Omega_0 \int_0^x \frac{\mathcal{F}'(\underline{x}')}{a(x')} \frac{x'(x-x')}{x} \frac{\sin\left[(\Omega_0-1)^{1/2}(x-x')\right]}{(\Omega_0-1)^{1/2}(x-x')} dx' \quad \text{for } \Omega_0 > 1 \\
&= -3 \int_0^x \frac{\mathcal{F}'(\underline{x}')}{a(x')} \frac{x'(x-x')}{x} dx' \quad \text{for } \Omega_0 = 1 \\
&= -3\Omega_0 \int_0^x \frac{\mathcal{F}'(\underline{x}')}{a(x')} \frac{x'(x-x')}{x} \frac{\sinh\left[(1-\Omega_0)^{1/2}(x-x')\right]}{(1-\Omega_0)^{1/2}(x-x')} dx' \quad \text{for } \Omega_0 < 1
\end{aligned} \tag{23}$$

This can be written in a more general way

$$-\Delta M = -3\Omega_0 \int_0^x \frac{\delta(\underline{x}')}{a(x')} x' \left(1 - \frac{x'}{x}\right) j_0\left[(\Omega_0-1)^{1/2}(x-x')\right] dx' \tag{24}$$

$$p = -3\Omega_0 \int_0^x \frac{\mathcal{F}'(\underline{x}')}{a(x')} x' \left(1 - \frac{x'}{x}\right) j_0\left[(\Omega_0-1)^{1/2}(x-x')\right] dx' \tag{25}$$

This is the general solution for a single source at distance $x$. Here $j_0(y) = \sin(y)/y$ and $j_0(0) = 1$. Remember that $j_0(iy) = \sinh(y)/y$.

### 2.3 Calculation for a Distribution of Sources

In a normal observational situation, the sources are not all at the same distance. We take a distribution of sources $n(x)$ so that

$$\int_0^{2/\Omega_0} dx \, n(x) = 1 \tag{26}$$

The upper bound is given by the horizon distance. A potential fluctuation at distance $x'$ will affect only sources at larger distance $x > x'$. The magnification and polarization measured at a given position on the sky is the mean of the polarizations of all the sources at different distances at that position. So a density perturbation at distance $x$ induces a polarization for all more distant sources at the same position in the sky. Thus the observed magnification $\Delta M = -2\delta x_1/x$ and $p = \delta x_2/x$ are

$$-\Delta M = -3\Omega_0 \int_0^{2/\Omega_0} dx' \frac{\delta(\underline{x}')}{a(x')} x' w\left(x', \Omega_0\right) \tag{27}$$

$$p = -3\Omega_0 \int_0^{2/\Omega_0} dx' \frac{\mathcal{F}'(\underline{x}')}{a(x')} x' w\left(x', \Omega_0\right) \tag{28}$$

$$\begin{aligned}
w\left(x'\right) &= \int_{x'}^{2/\Omega_0} dx \, n(x) \left(1 - \frac{x'}{x}\right) \frac{\sin\left((\Omega_0-1)^{1/2}(x-x')\right)}{(\Omega_0-1)^{1/2}(x-x')} \quad \text{for } \Omega_0 > 1 \\
&= \int_{x'}^{2} dx \, n(x) \left(1 - \frac{x'}{x}\right) \quad \text{for } \Omega_0 = 1
\end{aligned}$$

$$= \int_{x'}^{2/\Omega_0} dx\, n(x) \left(1 - \frac{x'}{x}\right) \frac{\sinh\left((1-\Omega_0)^{1/2}(x-x')\right)}{(1-\Omega_0)^{1/2}(x-x')} \quad \text{for } \Omega_0 < 1 \tag{29}$$

or written in a more general fashion

$$w(x', \Omega_0) = \int_{x'}^{2/\Omega_0} dx\, n(x) \left(1 - \frac{x'}{x}\right) j_0\left((\Omega_0 - 1)^{1/2}(x - x')\right) \tag{30}$$

This is the general solution for an arbitrary distribution of sources. This is useful if we know, or assume, a mass distribution of scatterers. It is also the basis for a statistical distribution of scatterers, ie. if we assume a cosmological model.

## 3 CALCULATING STATISTICAL PROPERTIES

### 3.1 Polarization Correlation Function

I now look at the statistical properties of the polarization field on the sky and I do the general case with a distribution of sources. The situation where all the sources are at the same distance is just a special case. First I calculate the two-point correlation function of the polarization and the magnification. I do a plane wave decomposition of the density field in the universe. The Fourier transform of the density field and of $\mathcal{F}'$ differ only by a phase factor, and in the plane-parallel approximation the two-point correlation functions are identical (B91 and M94). The plane wave decomposition in curved space is only meaningful for wavelengths much smaller than the radius of curvature. We also assume the plane-parallel approximation, i.e. that only waves nearly perpendicular to the beam are important. Then we can write

$$\int \frac{d^3k}{(2\pi)^3} \tilde{\delta}(\underline{k}) \exp(i\underline{k}\underline{x}) = \delta(\underline{x})$$

$$\int \frac{d^3k}{(2\pi)^3} \tilde{\delta}(\underline{k}) \exp(2i\psi) \exp(i\underline{k}\underline{x}) = \mathcal{F}'(\underline{x}) \; ; \; \tan(\psi) = \frac{k_2}{k_1} \tag{31}$$

In equations 27 and 28 we see that the expansion factor $a$ enters explicitly. The terms $\delta(\underline{x}')/a$ and $\mathcal{F}'(\underline{x}')/a$ are evaluated at the epoch at which the lightray is at that position. Both quantities are expected to evolve in the same way with $x$ since they are both second derivatives of the potential. In a flat universe where the growth of density fluctuations is well described by linear theory these terms are equal to $\delta_0(\underline{x}')$ and $\mathcal{F}'_0(\underline{x}')$ which are just the values at the current epoch, i.e. $a = 1$. For a non-flat universe the situation is, as usual, not so simple. We set

$$\frac{\delta(\underline{x})}{a} \equiv f(\underline{x}(a), \Omega_0) \times \delta_0(\underline{x}) \; ; \; \frac{\mathcal{F}'(\underline{x})}{a} \equiv f(\underline{x}(a), \Omega_0) \times \mathcal{F}'_0(\underline{x}) \tag{32}$$

In general, $f$ will be a function of position but we are going to assume that the large scale structure evolves simply by increasing the density contrast so $f = f(a, \Omega_0)$. Notice that this parametrization does not require linear growth of density fluctuations. The quantity $f$ will be discussed in more detail in §3.5 .

We first calculate the two-point correlation of the polarization $C_{pp}(\theta)$. This is the same as the magnification two-point correlation function. The Fourier Transforms of the magnification and polarization differ only by a phase factor. The amplitudes are the same and thus the power spectra and two point correlation functions will be identical. Two direction vectors $\hat{\underline{x}}$ and $\hat{\underline{x}}'$ are separated by an angle $\theta \ll 1$, $\underline{x} = \hat{\underline{x}}x$, and $\underline{x}' = \hat{\underline{x}}'x'$. $C_{pp}(\theta)$ can then be calculated as the average over the sky of the product of the polarizations measured in directions separated by distance $\theta$. For simplicity we let one direction vector be the z-axis.

$$\begin{aligned}
C_{pp}(\theta) &= \left\langle \Delta M(\hat{\underline{x}}) \Delta M(\hat{\underline{x}}') \right\rangle = \left\langle p(\hat{\underline{x}}) p^*(\hat{\underline{x}}') \right\rangle \\
&= \left\langle 9\Omega_0^2 \int_0^{2/\Omega_0} dx\, f(x)\, x\, w(x) \delta_0(\underline{x}) \int_0^{2/\Omega_0} dx'\, f'(x')\, x'\, w(x') \delta_0(\underline{x}') \right\rangle \\
&= 9\Omega_0^2 \int_0^{2/\Omega_0} dx\, f(x)\, x\, w(x) \int \frac{d^3k}{(2\pi)^3} \tilde{\delta}_0(\underline{k}) \exp(i\underline{k}\underline{x}) \times \\
&\quad \int_0^{2/\Omega_0} dx'\, f'(x')\, x'\, w(x') \int \frac{d^3k}{(2\pi)^3} \tilde{\delta}_0^*(\underline{k}') \exp(-i\underline{k}\underline{x}') \\
&= 9\Omega_0^2 \int_0^{2/\Omega_0} dx\, f(x)\, x\, w(x) \int_0^{2/\Omega_0} dx'\, f'(x')\, x'\, w(x') \int d^3k P(k) \exp(i(\underline{k}\underline{x} - \underline{k}\underline{x}'))
\end{aligned} \tag{33}$$

We have used the definition of $P(k)$, which is evaluated at the present epoch, so that

$$\langle \tilde{\delta}_0(\underline{k})\tilde{\delta}_0^*(\underline{k}')\rangle = (2\pi)^6 P(k)\delta(\underline{k}-\underline{k}') \tag{34}$$

We now need to use the small angle approximation $\theta \ll 1$ and the plane-parallel approximation, that only waves nearly perpendicular to the line-of-sight contribute to the polarization. This enters in the $\mu$ integral where the limits of integration are extended to $\pm\infty$. I set

$$\underline{k}=k(\cos\phi,\sin\phi,\mu) \; ; \; \underline{x}=x(0,0,1) \; ; \; \underline{x}'=x'(\theta,0,1) \; ; \; \underline{x}-\underline{x}'=(-x'\theta,0,x-x') \tag{35}$$

$$\begin{aligned}
&\int d^3k P(k)\exp(i(\underline{k}\underline{x}-\underline{k}\underline{x}'))\\
=&\int_0^\infty dk k^2 P(k)\int_0^{2\pi}d\phi\exp(-ik\cos\phi x'\theta)\int_{-\infty}^\infty d\mu\exp(ik\mu(x-x'))\\
=&\,\delta(x-x')4\pi^2\int_0^\infty dk k P(k)J_0(kx\theta)
\end{aligned} \tag{36}$$

We insert this result into the equation for $c_{pp}$.

$$C_{pp}(\theta)=36\Omega_0^2\pi^2\int_0^{2/\Omega_0}dx\, f^2(x)x^2w^2(x)\int_0^\infty dk k P(k)J_0(kx\theta) \tag{37}$$

We see that the two-point correlation function is a first moment of the power spectrum of density fluctuations. For a source at distance $x$, only waves with wavenumber less than $(x\theta)^{-1}$ contribute significantly to the correlation function.

We wish to relate the two-point polarization correlation function to the two-point correlation function of the mass $\xi(r)$. For this we can use the relationship between $P(k)$ and $\xi(r)$, they are Fourier Transforms of each other.

$$P(k)=\frac{1}{2\pi^2}\int_0^\infty dy\, y^2\xi(y)j_0(ky) \tag{38}$$

When we insert this relationship we get for $C_{pp}(\theta)$ that

$$\begin{aligned}
C_{pp}(\theta) &= 18\Omega_0^2\int_0^{2/\Omega_0}dx\, f^2(x)x^2w^2(x)\int_0^\infty dy\, y^2\xi(y)\int_0^\infty dk k J_0(kx\theta)j_0(ky)\\
&= 18\Omega_0^2\int_0^{2/\Omega_0}dx\, f^2(x)x^2w^2(x)\int_{x\theta}^\infty dy\, y\xi(y)\left(y^2-(x\theta)^2\right)^{-1/2}\\
&= 18\Omega_0^2\int_0^{2/\Omega_0}dx\, f^2(x)x^2w^2(x)\int_0^\infty dy\,\xi\left(\sqrt{(x\theta)^2+y^2}\right)\\
&= 9\Omega_0^2\int_0^{2/\Omega_0}dx\, f^2(x)x^2w^2(x)W(x\theta)
\end{aligned} \tag{39, 40}$$

For a source of distance $x$, the contribution to $C_{pp}(\theta)$ can be written as an integral over the correlation function on scales larger than $x\theta$. The quantity $W(x\theta)$ is the two-point correlation function of the surface density distribution evaluated at angle $\theta$. As a simple example of what might be observed is seen in Figure 1 I show $[C_{pp}(\theta)]^{1/2}$ for a CDM and an HDM power spectrum, Bardeen et al. 1986, where the powerspectra are normalized so that the rms density fluctuation in a sphere of radius $8h^{-1}$Mpc for $h$=1/2, equals unity. For simplicity it is assumed that all sources are at redshift $z$=3/4 and $\Omega_0=1$.

The lower curves (CDM solid, HDM dotted) are the curves for the correlation function on a scale up to 20 arcmins. An angular scale of 10 arcmins corresponds to a linear scale of 4.2 $h^{-1}$ Mpc at that redshift. On that angular scale both CDM and HDM have a correlation amplitude of $(0.02)^2$. On smaller scales CDM greatly increases that amplitude while for HDM the amplitude stays constant.

### 3.2   Polarization Power Spectrum

Let us look at the power spectrum of polarisation fluctuations $Q(\varpi)$. The power spectrum of polarization fluctuations is the Fourier transform of the two-point correlation function $C_{pp}(\theta)$. Notice that this definition of $Q$ is different from the definition in M94 and Gould & Villumsen 1994. In those papers $Q=Q(k)$ where k is an inverse length while in this paper $Q=Q(\varpi)$, $\varpi$ is an inverse angle. If the all the sources are at distance $x$ the two definitions differ by a factor $x^2$.

$$Q(\varpi) = \frac{1}{2\pi}\int_0^\infty d\theta\,\theta C_{pp}(\theta)J_0(\varpi\theta)$$

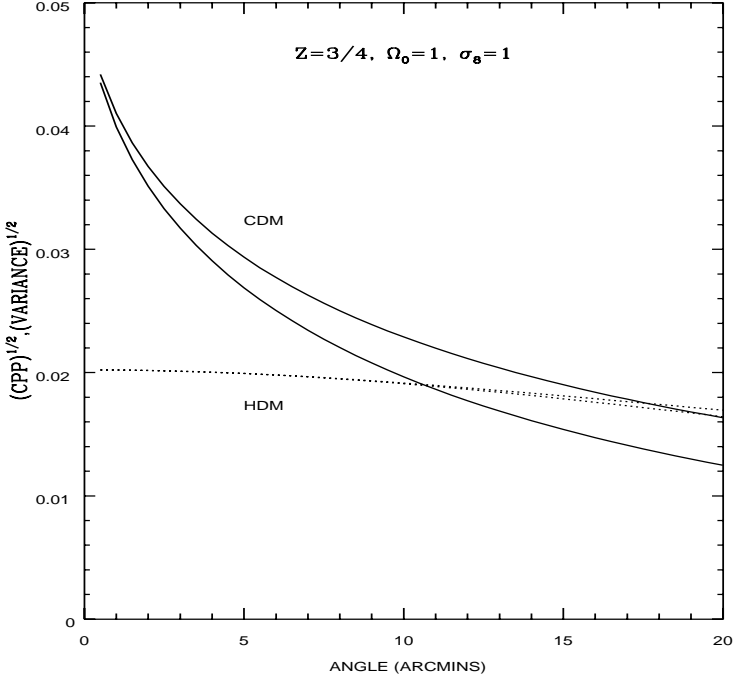

**Figure 1.** Square root of polarization correlation function $C_{pp}(\theta)$ and rms polarization fluctuations for CDM (solid lines), and HDM (dotted lines). All sources at $z = 3/4$, $\Omega_0 = 1$. Upper curves are rms fluctuations, lower curves are the square root of correlation amplitude.

$$\begin{aligned}
&= 18\pi\Omega_0^2 \int_0^{2/\Omega_0} dx\, f^2(x)w^2(x)x^2 \int_0^\infty dk k P(k) \int_0^\infty d\theta\, \theta J_0(kx\theta)J_0(\varpi\theta) \\
&= 18\pi\Omega_0^2 \int_0^{2/\Omega_0} dx\, f^2(x)w^2(x)x^2 \int_0^\infty dk k P(k) \frac{\delta(k-\varpi/x)}{x^2 k} \\
&= 18\pi\Omega_0^2 \int_0^{2/\Omega_0} dx\, f^2(x)w^2(x) P(\varpi/x) \qquad (41)
\end{aligned}$$

Notice that in this derivation I have used $\delta$ as symbol for the Dirac delta function. We can use the power spectrum of potential fluctuations $P_\Phi(k)$ which is related to $P(k)$

$$\nabla^2 \Phi(\underline{x}) = 4\pi G \delta(\underline{x}) \;\; \Rightarrow \tilde{\Phi}(\underline{k}) = -\frac{4\pi G}{k^2}\tilde{\delta}(\underline{k}) \;\; \Rightarrow P_\Phi(k) = 16\pi^2 G^2 \frac{P(k)}{k^4} \Rightarrow \qquad (42)$$

$$\frac{Q(\varpi)}{\varpi} = 18\pi\Omega_0^2 \int_{\frac{\varpi\Omega_0}{2}}^\infty dk f^2(\varpi/k) w^2(\varpi/k) P(k) k^{-2} \qquad (43)$$

$$= \frac{9\Omega_0^2}{8\pi G^2} \int_{\frac{\varpi\Omega_0}{2}}^\infty dk f^2(\varpi/k) w^2(\varpi/k) P_\Phi(k) k^2 \qquad (44)$$

This equation says that the contribution to the power spectrum of polarization fluctuations at inverse angle $\varpi$ from structure at a distance $x$ comes from the power spectrum of density fluctuations at wavenumbers larger than $\varpi/x$.

In Figure 2 I show the standard power spectra of density fluctuations $P(k)k^2$ for CDM and HDM, Bardeen *et al.* 1986 (Solid lines). The normalisations of amplitude are arbitrary. On the same plot is shown $Q(\varpi)\varpi$ for all sources at distance $x = 1$,ie. $z = 3$ (Dotted lines). This is done in order to directly compare the powerspectra for density fluctuations and for polarization fluctuations. In this case $\varpi = k$. I have multiplied $P(k)$ by two powers of the wavenumber and $Q(k = \varpi)$ by one power to demonstrate on what scales the density fluctuations and polarization fluctuations are generated. We see that the polarization fluctuations are generated at significantly larger scales than the density fluctuations. The range of scales that contribute to the polarization is also larger.

We can get some insight into equations 43 and 44 by considering a simple calculation for a flat universe. Suppose all the sources are at distance $x_0$ and $P(k)$ can be locally approximated by a power law of slope $n$ then we have approximately that

$$\frac{Q(\varpi)}{\varpi} \propto \int_{\frac{\varpi}{x_0}}^{\infty} dk \left(1 - \frac{\varpi}{kx_0}\right)^2 k^{n-2} \qquad (45)$$

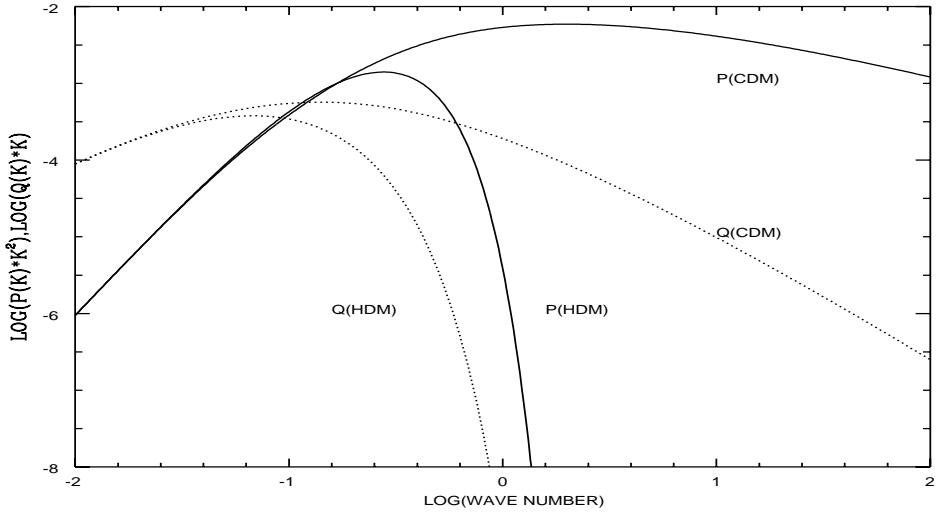

**Figure 2.** Power spectra of density fluctuations $P(k)k^2$, and polarization fluctuations $Q(\varpi = k)k$ for CDM and HDM for sources at distance $x = 1$. All absolute amplitudes are arbitrary.

This integrand peaks at

$$\frac{\varpi}{x_p} \equiv k_p = \left(\frac{4-n}{2-n}\right) \frac{\varpi}{x_0} \Rightarrow x_p = \left(\frac{2-n}{4-n}\right) x_0 \qquad (46)$$

This says that the peak contribution to the polarization occurs at distance $x_p$ from the observer. Let us consider CDM as an example. If we are looking on the Harrison-Zeldovich tail, $n = 1$, then $x_p/x_0 = 1/3$ meaning scales closer to the observer than to the source are most effective. If we look at galaxy scales, $n \sim -2$, then $x_p/x_0 = 2/3$, most of the power is generated nearer the source. For positive slope of $P(k)$, more of the power is on smaller scale, which pushes the most effective lens distance towards the observer. For a negative slope of $P(k)$, the opposite is true.

How meaningful is the simple $f$ parametrization? We see from Eq. 44 that the power spectrum $Q$ can be derived as an integral over the power spectrum of potential fluctuations with a high pass filter $w$. Potential fluctuations are weighted towards large scales so the simple parametrization is expected to work well. Notice again, that this parametrization does not require linear theory.

### 3.3 Polarization Variance

Another important statistical quantity is the polarization variance on a scale $\theta$ using an axially symmetric window function $W_1(y/\theta)$, where

$$2\pi \int_0^{\infty} dy\, y W_1(y/\theta) = 1 \qquad (47)$$

$$\tilde{W}_1(\varpi\theta) = \int \frac{dy}{2\pi} y W_1(y/\theta) J_0(y\varpi) \qquad (48)$$

The variance of a homogeneous field on a scale $\theta$ is the mean square amplitude of the field when convolved with a filter function $W_1(y/\theta)$. This variance can be evaluated simply as

$$\sigma_p^2(\theta) = 2\pi \int_0^\infty d\varpi\, \varpi\, Q(\varpi) \tilde{W}_1^2(\varpi\theta)$$

$$= 36\Omega_0^2 \pi^2 \int_0^{2/\Omega_0} dx\, f^2(x) x^2 w^2(x) \int_0^\infty dk\, k\, P(k) \tilde{W}_1^2(kx\theta) \tag{49}$$

where we have used Eq. 41 and that $k \cdot x = \varpi$. For the special cases of a tophat filter and a Gaussian filter of radius $\theta$ we get

$$\sigma_p^2(\theta) = 36\Omega_0^2 \pi^2 \int_0^{2/\Omega_0} dx\, f^2(x) x^2 w^2(x) \int_0^\infty dk\, k\, P(k) \left(\frac{2J_1(kx\theta)}{kx\theta}\right)^2 \tag{50}$$

$$\sigma_p^2(\theta) = 36\Omega_0^2 \pi^2 \int_0^{2/\Omega_0} dx\, f^2(x) x^2 w^2(x) \int_0^\infty dk\, k\, P(k) \exp(-k^2 x^2 \theta^2) \tag{51}$$

The inner integral in Eqs. 49, 50, and 51 is just $\pi/2$ times the mean square density fluctuations $\sigma_\mu^2$ on a sheet perpendicular to the line of sight with the appropriate window function. Notice that $\sigma_\mu$ is the volume density fluctuations evaluated on a sheet, *not* the surface density fluctuations. Thus we get that

$$\sigma_p^2(\theta) = 18\Omega_0^2 \pi^3 \int_0^{2/\Omega_0} dx\, f^2(x) x^2 w^2(x) \sigma_\mu^2(x\theta) \tag{52}$$

Neither the two-point correlation function $C_{pp}(\theta)$ nor the variance $\sigma_p^2$ depend on the statistics of the polarization field. Figure 1 also shows $\sigma_p$ for CDM and HDM for the same parameters as in §3.1(upper curves). The filter function is a tophat. On a scale of 5 arcmin, which is the scale measured by M94, the CDM prediction is an rms amplitude of 3% while HDM predicts 2% for $\Omega_0 = 1$.

### 3.4   Gaussian Polarization Statistics

If the field of density fluctuations is a Gaussian random field then the polarization field will also be a Gaussian random field. Polarization results from scattering of lightrays off the density fluctuations. If the density field is not Gaussian then if the scatterers are uncorrelated, by the Central Limit Theorem, the polarization field will become Gaussian for a sufficiently large pathlength. We know that the density field is correlated but if the pathlength is much larger than the correlation length of the density field the Central Limit Theorem will still imply that the polarization field will be Gaussian.

In that case $Q(\varpi)$ fully specifies the statistics of the polarization field. Eq. 41, or equivalently Eqs. 43, 44, is the fundamental equation for the statistical study of weak lensing. Given a known source distribution, then from the power spectrum of density fluctuations and $\Omega_0$ we can predict the polarization field. The fact that the polarization field is Gaussian means there will be specific predictions for the higher order correlations of the field given the two-point correlation. The distribution of polarizations will be Gaussian. If we have observed a polarization field, i.e. $<\chi>$, as a function of position on the sky, then if this field does not have Gaussian statistics or the higher order correlations do not agree with predictions based on the two-point correlation function, then we can falsify the weak lensing hypothesis for this $<\chi>$ field. There is a caveat to this statement, single scattering events from individual clusters can be stronger than the polarization in the field. Thus in the tail of the distribution we can be dominated by single scattering events and the polarization distribution need then not be Gaussian.

If the polarization field is a complex Gaussian field, then the amplitude field will have a Rayleigh distribution. In that case, the most likely measured amplitude will be the rms amplitude. Thus for M94 the most likely polarization amplitude in an unbiased, flat CDM universe will be 3%.

### 3.5   Estimate of Curvature Effects

Let us look at the effects of varying $\Omega_0$. For simplicity we put all sources at distance $x$, or equivalently redshift $z$. This makes the weightfunction $w$ simple

$$w(x', \Omega_0) = \left(1 - \frac{x'}{x}\right) j_0\left((\Omega_0 - 1)^{1/2}(x - x')\right) \tag{53}$$

We look at the rms polarization $\sigma_p(0)$ as a function of $\Omega_0$ for a given source redshift $z$. At low redshift, $z \ll 1$, we have $x = z$ for any value of $\Omega_0$ and the weight function $w$ is independent of $\Omega_0$. That means $\sigma_p(0)$ is proportional to $\Omega_0$ for a fixed source redshift. This proportionality makes physical sense since then the polarization is proportional to the potential well depth. If we go to higher values of $z$ three effects enter that all pull in the same direction. For a given $z$, as we lower $\Omega_0$, the distance increases, the weight function $w$ increases, and $f$ increases. The distance is larger because normalised to the present epoch the universe is older and therefore the lightray has travelled further. Secondly, $w$ has increased because $x$ has increased and the lightrays are more divergent (for $\Omega_0 < 1$) or less convergent (for $\Omega_0 > 1$). The different cosmological models also enter in the terms $\delta(\underline{x}')/a$ and $\mathcal{F}'(\underline{x}')/a$ evaluated at the epoch at which the lightray is at that position. Both quantities are expected to evolve in the same way with $x$ since they are both second derivatives of the potential. In a flat universe where the growth of density fluctuations on relevant scales is well described by linear theory these terms are just equal to $\delta_0(\underline{x}')$ and $\mathcal{F}'_0(\underline{x}')$ which are just the values at the current epoch, i.e. $a = 1$. For a non-flat universe the equations for $f$ are more complicated, however the validity of the linear theory estimate of $f$ is not affected. In linear theory (Peebles 1980) the amplitude of the growing mode, and therefore $f$, can be calculated for any value of $\Omega_0$. We can get a rough estimate of $f$ from linear theory where the logarithmic growth rate of the growing mode is approximately $\Omega_0^{0.6}$. Thus we get

$$f \approx a^{(\Omega_0^{0.6}-1)} \approx (1-z)^{(\Omega_0^{0.6}-1)} \approx \left(1 + \left[1 - \Omega_0^{0.6}\right]z\right) \text{ for } z \ll 1 \tag{54}$$

We see from this simple parametrisation that $f$ decreases with $\Omega_0$ and increases rapidly with distance. This effect is the most important of the three effects.

B91 indicate from N-body simulation for Cold Dark Matter and Hot Dark Matter that for the purposes of weak lensing calculations it is a good approximation to use linear theory for the growth of density fluctuations. There is, however, a finite probability that light rays will pass close to a dense cluster where the deflections will be strong. In this case the formalism will break down and we need to talk about strong lensing. However, the fraction of sky covered with strong lenses is small and this problem will be ignored in the rest of the paper.

From Eq 52 we see that the amplitude of the curvature effects is independent of $P(k)$. We can thus show simply the curvature correction $C$, which we define as the rms polarization fluctuations at zero separation as a function of $\Omega_0$ and the redshift of the sources $z$ measured relative to the rms fluctuations for $\Omega_0 = 1$. Here we have for illustrative purposes assumed that all the source are at the same redshift.

$$C(\Omega_0, z) \equiv \frac{\sigma_p(0; z; \Omega_0)}{\sigma_p(0; z; \Omega_0 = 1)} \tag{55}$$

The curvature correction can be interpreted as the value of the density parameter that we would infer from observations if we naively assumed that the universe were flat ($\Omega_0 = 1$), and that the growth of density fluctuations were linear in the expansion factor $a$.

In Figure 3 we show the curvature correction $C(\Omega_0, z)$ as a function of $z$ for fixed values of $\Omega_0$ ranging from a very open universe ($\Omega_0 = 0.2$) to a very closed universe ($\Omega_0 = 2.0$). At low redshifts we see that $C(\Omega_0, z \ll 1) \approx \Omega_0$ but at higher redshift there are significant extra curvature effects. For open universes the curvature correction is an increasing function of source redshift $z$ while $C$ decreases with $z$ for a closed universe. In other words, lensing amplitudes will approach the values for a flat universe. A simple way of looking at this is to see that as we go to higher redshifts, $\Omega(z, \Omega_0)$ will approach unity. This is demonstrated by the dotted curves where we have plotted $\Omega(\Omega_0, z)$ on top of the curvature correction for the same values of $\Omega_0$ in Figure 3. For a given value of $\Omega_0$ these curves lie nearly on top of each other showing that what we are really measuring is the value of $\Omega$ at the epoch when the photons were emitted from the source. In other words

$$C(\Omega_0, z) \approx \Omega(\Omega_0, z) = \frac{\Omega_0 + \Omega_0 z}{1 + \Omega_0 z} \tag{56}$$

That means if we know the source redshift and the value of $\Omega$ at that epoch we can infer the current value $\Omega_0$ from the equation.

$$\Omega_0 = \frac{\Omega(z)}{1 + (1 - \Omega(z))z} \tag{57}$$

We can thus write an approximate equation for the rms polarization fluctuations in terms of source redshift $z$.

$$\sigma_p^2(0) = C_{pp}(0) \approx 36\pi^2 \left(2\left[1 - (1+z)^{-1/2}\right]\right)^3 \Omega_0^2 \left(\frac{1+z}{1+\Omega_0 z}\right)^2 \int_0^\infty dk k P(k) \text{ for } z \lesssim 1 \tag{58}$$

$$\sigma_p^2(0) = C_{pp}(0) \approx 36\pi^2 z^3 (1 + 2(1-\Omega_0)z)\Omega_0^2 \int_0^\infty dk k P(k) \text{ for } z \ll 1 \tag{59}$$

In the same way we can write $C_{pp}(\theta)$ and $\sigma_p^2(\theta)$ in terms of the source redshift $z$ for a given value of $\Omega_0$. Since $x(\Omega_0, z)$ is a decreasing function of $\Omega_0$ for fixed $z$ we find that as we lower $\Omega_0$, both $C_{pp}(\theta)$ and $\sigma_p^2(\theta)$ will be steeper functions of $\theta$.

$$C_{pp}(\theta, z, \Omega_0) \approx \Omega_0^2 \left(\frac{1+z}{1+\Omega_0 z}\right)^2 C_{pp}(\theta \cdot x(\Omega_0, z)/x(\Omega_0 = 1, z), z, \Omega_0 = 1), \ z \lesssim 1 \tag{60}$$

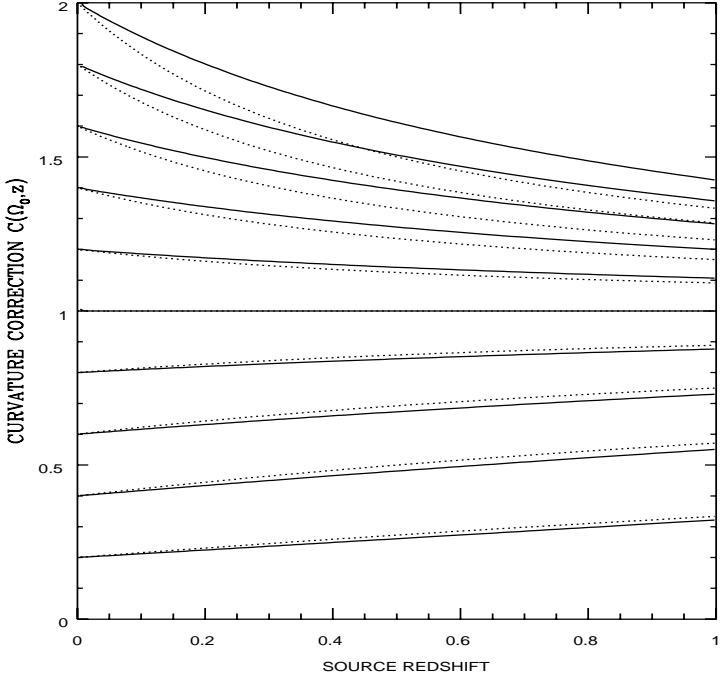

**Figure 3.** Curvature corrections for rms polarization fluctuations. Solid lines shows $C(\Omega_0, z)$ at different source redshifts for a set of values of $\Omega_0$. The dotted lines show the value of $\Omega$ at the source redshift.

$$\sigma_p^2(\theta, z, \Omega_0) \approx \Omega_0^2 \left(\frac{1+z}{1+\Omega_0 z}\right)^2 \sigma_p^2(\theta \cdot x(\Omega_0, z)/x(\Omega_0 = 1, z), z, \Omega_0 = 1), \ z \lesssim 1 \tag{61}$$

The weak lensing prediction for a zero cosmological constant universe is that observations of the polarization amplitude as a function of source redshift will follow a curve like those shown in Figure 3. If the amplitude does not follow such a curve, it is an indication for non-zero $\Lambda$.

## 4 SUMMARY OF RESULTS

In a universe with an FRW metric, defined by the current cosmological density parameter $\Omega_0$, and a zero cosmological constant, we have a density field $\delta_0(\underline{x})$, defined at the current epoch with a power spectrum $P(k)$. It is assumed that the growth of density fluctuations can be simply described by a universal function $f(x, \Omega_0)$. It is not necessary to assume linear growth factors. The shear in the gravitational field is desribed by the terms $\mathcal{F}'$ defined in equation 6. The distribution of sources in comoving angular diameter distance $x$ is $n(x)$. Other terms are the expansion $a$, the inverse angle $\varpi$, the power spectrum of potential fluctuation $P_\Phi$, and the variance of the volume density on a sheet, $W$.

$$f \equiv \frac{\delta(\underline{x})}{\delta_0(\underline{x})a} \equiv \frac{\mathcal{F}'(\underline{x})}{\mathcal{F}'_0(\underline{x})a} \tag{62}$$

$$w(x', \Omega_0) = \int_{x'}^{2/\Omega_0} dx\, n(x) \left(1 - \frac{x'}{x}\right) j_0\left((\Omega_0 - 1)^{1/2}(x - x')\right) \tag{63}$$

$$-\Delta M = -3\Omega_0 \int_0^{2/\Omega_0} dx' \frac{\delta(\underline{x}')}{a(x')} x' w(x', \Omega_0) \tag{64}$$

$$p = -3\Omega_0 \int_0^{2/\Omega_0} dx' \frac{\mathcal{F}'(\underline{x}')}{a(x')} x' w(x', \Omega_0) \tag{65}$$

$$C_{pp}(\theta) = 36\Omega_0^2 \pi^2 \int_0^{2/\Omega_0} dx\, f^2(x) x^2 w^2(x) \int_0^\infty dk\, k P(k) J_0(kx\theta) \tag{66}$$

$$= 9\Omega_0^2 \int_0^{2/\Omega_0} dx \ f^2(x) x^2 w^2(x) W(x\theta) \tag{67}$$

$$Q(\varpi) = 18\pi\Omega_0^2 \int_0^{2/\Omega_0} dx \ f^2(x) w^2(x) P(\varpi/x) \tag{68}$$

$$= \frac{9\Omega_0^2 \varpi}{8\pi G^2} \int_{\frac{\varpi \Omega_0}{2}}^{\infty} dk f^2(\varpi/k) w^2(\varpi/k) P_\Phi(k) k^2 \tag{69}$$

$$\sigma_p^2(\theta) = 36\Omega_0^2 \pi^2 \int_0^{2/\Omega_0} dx \ f^2(x) x^2 w^2(x) \int_0^\infty dk k P(k) \tilde{W}_1^2(kx\theta) \tag{70}$$

$$= 36\Omega_0^2 \pi^2 \int_0^{2/\Omega_0} dx \ f^2(x) x^2 w^2(x) \int_0^\infty dk k P(k) \left(\frac{2 J_1(kx\theta)}{kx\theta}\right)^2 \ \text{Tophat} \tag{71}$$

$$= 36\Omega_0^2 \pi^2 \int_0^{2/\Omega_0} dx \ f^2(x) x^2 w^2(x) \int_0^\infty dk k P(k) \exp(-k^2 x^2 \theta^2) \ \text{Gaussian} \tag{72}$$

If all the sources are the same redshift $z$ we can write the results in terms of redshift, where $C$ is the curvature correction.

$$\sigma_p^2(0) = C_{pp}(0) \approx 36\pi^2 \left(2\left[1 - (1+z)^{-1/2}\right]\right)^3 \Omega_0^2 \left(\frac{1+z}{1+\Omega_0 z}\right)^2 \int_0^\infty dk k P(k) \ \text{for } z \lesssim 1 \tag{73}$$

$$\sigma_p^2(0) = C_{pp}(0) \approx 36\pi^2 z^3 (1 + 2(1-\Omega_0)z)\Omega_0^2 \int_0^\infty dk k P(k) \ \text{for } z \ll 1 \tag{74}$$

$$C_{pp}(\theta, z, \Omega_0) \approx \Omega_0^2 \left(\frac{1+z}{1+\Omega_0 z}\right)^2 C_{pp}(\theta \cdot x(\Omega_0, z)/x(\Omega_0 = 1, z), z, \Omega_0 = 1), \ z \lesssim 1 \tag{75}$$

$$\sigma_p^2(\theta, z, \Omega_0) \approx \Omega_0^2 \left(\frac{1+z}{1+\Omega_0 z}\right)^2 \sigma_p^2(\theta \cdot x(\Omega_0, z)/x(\Omega_0 = 1, z), z, \Omega_0 = 1), \ z \lesssim 1 \tag{76}$$

$$C(\Omega_0, z) \approx \Omega(\Omega_0, z) = \frac{\Omega_0 + \Omega_0 \ z}{1 + \Omega_0 \ z} \tag{77}$$

## 5 DISCUSSION

The polarization field has been calculated rigorously, in the Born- approximation, for any value of the cosmological density parameter $\Omega_0$, but with zero cosmological constant. For a known mass distribution we calculate the polarization for a distribution of sources in real space. We show rigorously that at low redshift the polarization is proportional to $\Omega_0$. Thus weak lensing observations, such as in M94, is a direct measure of $\Omega_0$. This is in retrospect not surprising, since in that situation the scales of the universe that we are probing are much smaller than the radius of curvature of the universe. At higher redshifts explicit curvature effects manifest themselves in the calculation of $w(x)$ but only weakly. The main curvature effects are that the comoving angular distance $x$ is no longer equal to the redshift $z$ and that the growth of density fluctuations depends on $\Omega_0$. This formalism permits the comparison of weak lensing observations of large scale structure with theoretical models of the universe in an FRW metric.

In summary, the current formalism permits predictions of the polarization field and its statistical properties for a well defined cosmological model in an FRW universe with $\Lambda = 0$.

### Acknowledgments

I thank T. Brainerd, P. Schneider, R. van de Weygaert, and S. White for useful discussions, and I thank R. Blandford for teaching me how to do analytical work.